\documentclass[a4paper,12pt]{article}
\usepackage[dvipdfmx]{graphicx}
\usepackage{bm}
\usepackage{hyperref}
\usepackage{makeidx}
\usepackage{color}
\usepackage{here}
\usepackage{multirow}
\catcode`\@=11 
\def\lsim{\mathrel{\mathpalette\@versim<}}
\def\gsim{\mathrel{\mathpalette\@versim>}}
\def\@versim#1#2{\vcenter{\offinterlineskip
	\ialign{$\m@th#1\hfil##\hfil$\crcr#2\crcr\sim\crcr }}}
\catcode`\@=12 

\begin{document}
\begin{center}
{\Large
Operation of ILC250 at the Z-pole}\\
\vspace{10mm}
Kaoru Yokoya, Kiyoshi Kubo, and Toshiyuki Okugi, \\
High Energy Accelerator Research Organization (KEK), Japan \\
Aug.27. 2019
\end{center}

\small{
ILC (International Linear Collider) is under consideration as the next global 
project of particle physics. Its Technical Design Report, published in 2013, describes 
the accelerator for the center-of-mass energies above 200GeV. The operation of 
ILC at lower center-of-mass energies has not been studies intensively. 
This report discusses the operation of the ILC at a center-of-mass  91.2GeV 
and presents a possible parameter set.}

\section{Introduction}
When the serious design study of the ILC started in 2005, 
the first design criteria was 
``a continuous centre-of-mass energy range between 200 GeV and 500 GeV" 
(TDR\cite{bib:TDR}, page 3). 
Hence, the TDR quoted the luminosities only for the center-of-mass energies at 
200, 230, 250 GeV and above. Obviously, once ILC is built, lower energies 
such as Z-pole (91.2GeV) and W-pair threshold (160GeV) would be of interest, 
though these are not the main concern of ILC. 

In the baseline design of ILC the positron production scheme using undulator 
magnets is adopted. In this sheme the electron beam, before going to the 
collision point, goes through undulators to produce photons (over several MeV) 
which create positrons on a target. To this end the electron energy must be 
at least about 125 GeV. For operation at $E_{CM}\leq 250$ GeV TDR adopted the 
so-called 5+5Hz scheme\footnote{The electron linac is operated at 10Hz: 
5Hz to accelerate the beam to $\sim 150$ GeV which produces positrons and 
another 5Hz to accelerate the beam to $E_{CM}/2$ for collision experiment. 
This is sometimes referred to as `10Hz' operation. 
However, there is another 10Hz operation, in which all systems, the injectors, 
damping rings, main linacs etc., are operated at 10Hz to make 10Hz collisions. 
Thus, to distinguish from the latter we call the former `5+5Hz'.}

The possibilities of ILC operation at the Z-pole (and W-pair threshould) 
was first discussed by N.~Walker\cite{bib:NWalker}. Later, K.~Yokoya gave a short 
report at a workshop.\cite{bib:KYmorioka}. 
These reports gave a possible luminosity range at the Z-pole by using 
a scaling law and pointed out many challenges to be studied later.

The luminosity is given by
\begin{equation}
{\cal L} = \frac{f_{rep} n_{b} N^2}{4\pi \sigma^*_x \sigma^*_y} H_D
\end{equation}
where $f_{rep}$ is the repetition rate of the beam pulse, $n_{b}$ the 
number of bunches in a pulse, $N$ the number of particles per bunch, 
and $\sigma^*_{x(y)}$ is the horizontal (vertical) beam size at the 
IP (interaction point).  $H_D$ (luminosity enhancement factor)  expresses the effects 
of the beam-beam force. With the optics around the IP fixed, $\sigma^*_{x(y)}$ is proportional to the square root of the geometric emittance $\varepsilon_{x(y)}$. 
Since the geometric emittance is inversely proportional to the beam energy, 
a naive scaling expects ${\cal L} \propto E_{CM}$. 
However, the large geometric emittance at low energies causes a larger 
beam size at the final quadrupole magnets such that the beam halo may produce 
backgrounds to the experiments. Such halo particles usually are eliminated by 
collimators in upstream. However, too deep a collimation would cause 
further backgrounds and enhancement of beam fluctuation by the wake field 
in the collimators. In such a case one has to relax the beta functions at the IP, 
which brings about a lower luminosity.
If this happens in one of the $x$ and $y$ planes, then the luminosity would be 
proportional to $E_{CM}^{3/2}$, and if in both planes, to $E_{CM}^2$. 
The above report\cite{bib:NWalker} gave a luminosity range 
$1-1.5 \times 10^{33} /{\rm cm}^2/{\rm s}$, 
corresponding to ${\cal L} \propto E_{CM}^2$ or $E_{CM}^{3/2}$ based on the 
luminosities at 200-250GeV in the TDR.

\begin{table}[H]
\begin{center}
\caption\protect{\large Proposed Parameters for Operation at Z-pole  \\
{\small The baseline parameters for 250GeV is listed for reference.} \label{tab:param}}
\begin{tabular}{llrrl}
Center-of-Mass Energy &    $E_{CM}$     &   91.2  &  250  &  GeV \\
Beam Energy                &    $E_{beam}$   &  45.6  &  125  &  GeV   \\
Collision rate         &    $f_{col}$    &  3.7  &   5  &  Hz   \\
Electron linac repetition rate   &                &  3.7+3.7 &  5  &   Hz   \\
Pulse interval in electron main linac &       &   135  &  200   &    ms   \\
Electron energy for e$^+$ production     &    &   125  &  125  &   GeV   \\
Number of bunches      &    $n_b$                &  1312  &  1312  &  \\
Bunch population       &    $N$          &  2     &  2   &   $\times 10^{10}$   \\
Bunch separation       &  $\Delta t_b$      &  554   & 554 &   ns \\
RMS bunch length       &  $\sigma_z$         &  0.41 &  0.30  &   mm \\
Electron energy spread at IP (rms) &  $\sigma_{E_{-}}/E_{-}$     &  0.30 &  0.188 &   \%  \\
Positron energy spread at IP (rms) &  $\sigma_{E_{+}}/E_{+}$     &  0.30 &  0.150 &   \%  \\
Electron polarization  &   $P_-$         &   80  &  80 &     \%      \\
Positron polarization  &   $P_+$            &   30 &  30  &   \%    \\
Emittance from DR (x)  &  $\gamma \varepsilon^{DR}_x $   &   4  &  4  &  $\mu$m \\
Emittance from DR (y)  &  $\gamma \varepsilon^{DR}_y $      &   20 & 20 &    nm    \\
Emittance at the linac exit (x)    &  $\gamma \varepsilon^{IP}_x $   &   5 &  5 & $\mu$m   \\
Emittance at the linac exit (y)    &  $\gamma \varepsilon^{IP}_y $    &   35  & 30   &  nm   \\
Emittance at IP (x)    &  $\gamma \varepsilon^{IP}_x $   &   6.2 &  5 & $\mu$m   \\
Emittance at IP (y)    &  $\gamma \varepsilon^{IP}_y $    &   48.5  & 35   &  nm   \\
Beta x at IP           &   $\beta^*_x$     &  18   &   13  &  mm      \\
Beta y at IP           &   $\beta^*_y$     &  0.39  &  0.41  &   mm      \\
Beam size at IP (x)    &  $\sigma^*_x$    &  1.12   &  0.515 &  $\mu$m \\
Beam size at IP (y)    &  $\sigma^*_y$    &  14.6   & 7.66  &  nm    \\
Disruption Param (x)   &   $D_x$        &             0.41   &  0.51  &  \\
Disruption Param (y)   &   $D_y$        &             31.8  &  34.5  &   \\
Geometric luminosity   & ${\cal L}_{geo}$ &   0.95 & 5.29 & $10^{33}/{\rm cm}^2/{\rm s}$    \\
Luminosity             & ${\cal L}$       &   2.05  & 13.5 & $10^{33}/{\rm cm}^2/{\rm s}$   \\
Luminosity enhancement factor  &   $H_D$         &  2.16    &  2.55  &  \\
Luminosity at top 1\%   &                     &   99.0   &  74  & \%    \\
Number of beamstrahlung &   $n_\gamma$       &              0.841   & 1.91 &  \\
Beamstrahlung energy loss  &  $ \delta_{BS}$  &   0.157   & 2.62 &  \%  
\end{tabular}
\end{center}
\end{table}

Since the above studies there have been several changes in the ILC design. 
Most relevant ones for our context are:
\begin{itemize}
\setlength{\itemsep}{0mm}
\item  The initial center-of-mass energy for the ILC was reduced from 500GeV to 
250GeV in order to concentrate on the Higgs study.
\item  The undulator length was extended from 147m to 231m so that 
the electron beam of 125GeV can produce sufficient number of positrons.
Thus, we do not need 5+5Hz scheme for $E_{CM}=250$GeV.
\item  The normalized horizontal emittance at the IP was reduced  
from 10$\mu$m to 5$\mu$m by modifying the damping ring design. 
This change improved the luminosity at $E_{CM}=250$GeV by a factor $\sim 1.6$.
\cite{bib:AWLCSLAC}
\end{itemize}

In the present report, taking into account these changes we give study results 
on the possible luminosity at the Z-pole.

Table \ref{tab:param} shows the preliminary parameter set to be described 
in this report. The study results for each item will be discussed in the later sections.

\section{Repetition Rate \label{sec:RepRate}}
As mentioned in the introduction, the TDR adopted the `5+5Hz' scheme for the 
operation at low energies. This scheme demands more electric power than 
the normal 5Hz operation but this is possible because the power system of 
500GeV machine is available.  For example, the electron linac would produce 
45.6GeV and $\sim$150GeV electron beams alternatingly. The sum of the power 
of these two is less than the power for 250GeV electron beam.  

However, this is not true for the ILC 250GeV. The installed power 
is not sufficient for 5+5Hz operation. 
In this section we estimate the possible highest repetition rate 
under the limitation of the power system of ILC 250GeV.

First, let us discuss how to switch the beam energy.
To get a 45.6GeV electron beam with a 125GeV linac, one might want to accelerate 
the beam to  45.6GeV with the full gradient and turn off the power 
for the rest of the linac. This is better from the view point of the beam dynamics  
in the powered cavities. 
However, for avoiding the effects of the beam-induced field  
in the un-powered part of the linac, one has to detune the cavities quickly (a few Hz). 
This can only be done by piezo tuners but the dynamic range of the equipped tuners is 
far from this requirement. Thus we have to accelerate the electron beam uniformly 
over $\sim$5km with a very low gradient (8.76MV/m compared with 31.5MV/m for 125GeV)
\footnote{31.5$\times$(45.6-15)/(125-15)=8.76 MV/m,  
since the injection energy to the main linac is 15GeV.}
Thus, the electron linac must be operated with gradients 31.5MV/m and 
8.76MV/m, alternatingly. 

The possible parameters of the RF system are 
shown in Table \ref{tab:RFparam}. (This estimation was done by 
T.~Matsumoto.\cite{bib:Matsumoto}). 
The parameters of the normal operation for $E_{CM}=250$GeV are identical 
to the one in the left column except the repetition is 5Hz.
Since the gradient is low for the 45.6GeV beam, the klystron power is low and the 
fill time (and hence the RF pulse length) is short. 
One can find from this table that the integral of the modulator output 
in one cycle is 14.66MW$\times$1.65ms + 7.83MW$\times$1.06ms = 32.5kJ, 
which is larger than the first term (125GeV) by a factor  1.34. 
Therefore, the highest possible repetition rate is 5Hz/1.34 $\approx$ 3.7Hz.   
There will be some margin in the actually installed power which can raise the rate a little,  
but we adopt 3.7+3.7Hz operation in the following.

\begin{table}[H]
\caption{RF system parameters for alternating operation of 125GeV and 45.6GeV
\label{tab:RFparam}}
\begin{center}
\begin{tabular}{lrrl}
                                           & e$^+$ production  &  collision  &      \\
Final beam energy                    &      125     &   45.6   &  GeV   \\
Average accelerating gradient   &       31.5    &   8.76   & MV/m  \\
Peak power per cavity             &          189  &   77.2   &   kW   \\
Klystron peak power               &          9.82   &   4.15   &   MW   \\
Klystron efficiency                  &           67    &   53     &  \%    \\
Modulator output                   &        14.66    &   7.83    &  MW  \\
Fill time                               &         0.927    &   0.328   &   ms   \\
Beam pulse length                &          0.727    &   0.727  &    ms     \\
RF pulse length                    &           1.65     &   1.06    &    ms    \\
Repetition rate                     &            3.7     &    3.7     &   Hz
\end{tabular}
\end{center}
\end{table}

\section{Damping Rings}
Basically, the damping rings designed for the TDR are capable of 5+5Hz operation 
by a reinforcement of the wiggler magnets and the RF system. 
However, the design has been changed slightly since AWLC2017 at SLAC\cite{bib:AWLCSLAC} 
(smaller horizontal emittance $6\mu$m to $4\mu$m). 
It must be checked whether the new design accepts 3.7+3.7Hz operation. 

In 3.7+3.7Hz operation, the beam stay in the damping ring for 
1/3.7Hz/2=135ms, while 200ms in the baseline 5Hz operation. 

\begin{figure}[H]
\begin{center}
\includegraphics[width=0.6\textwidth, clip]{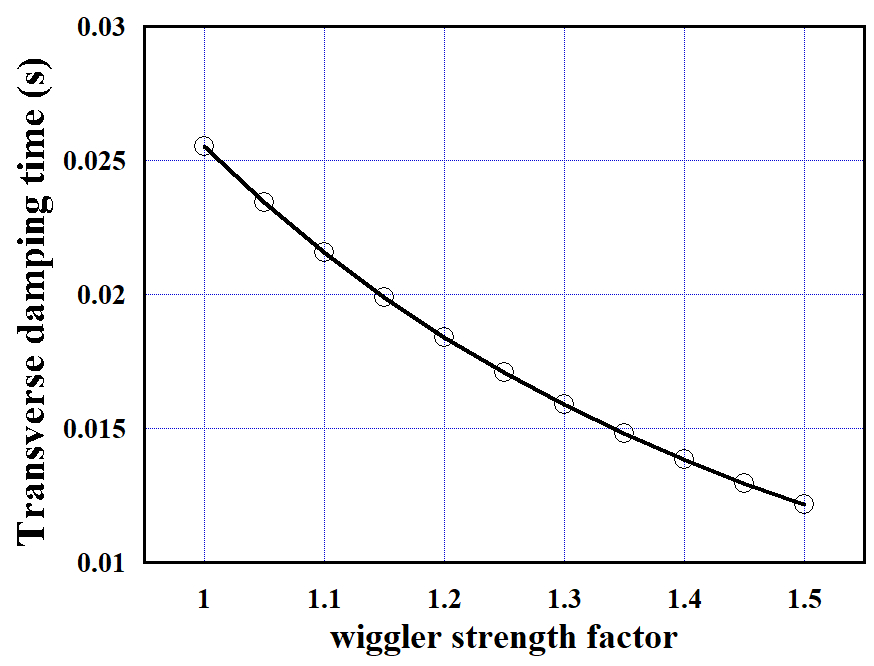}
\end{center}
\vspace{-10mm}
\caption{The transverse damping time as a function of the wiggler strength 
parameter. \label{fig:DampingTime}}
\end{figure}

Figure.\ref{fig:DampingTime} shows the transverse damping time as a function 
of the wiggler strength parameter. Factor 1 corresponds to the field 1.29T 
(curvature 0.07745m$^{-1}$).

\begin{figure}[H]
\begin{minipage}{0.49\textwidth}
\includegraphics[width=\textwidth, clip]{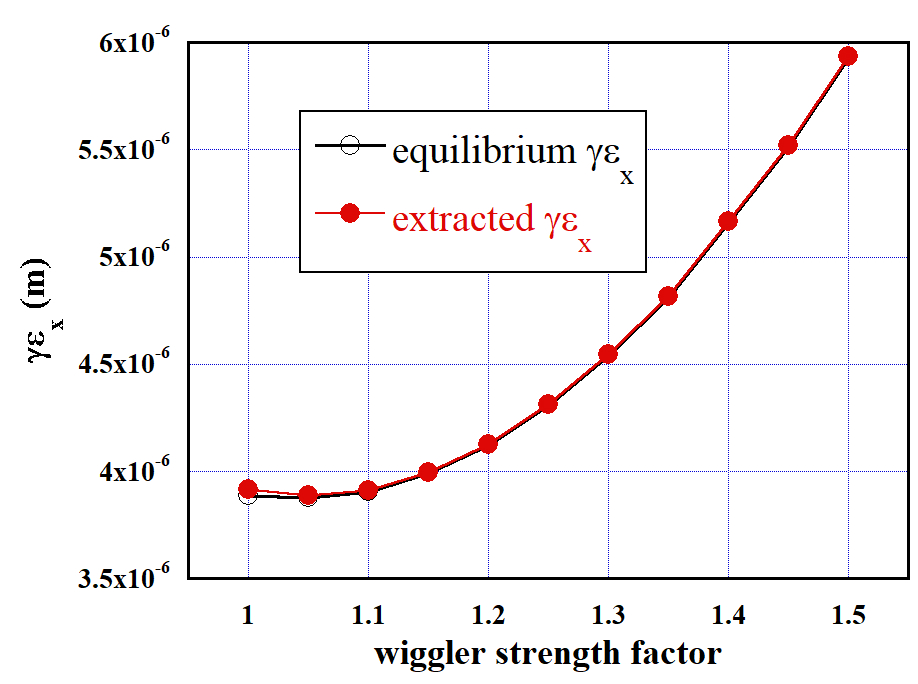}
\end{minipage}
\begin{minipage}{0.49\textwidth}
\includegraphics[width=\textwidth, clip]{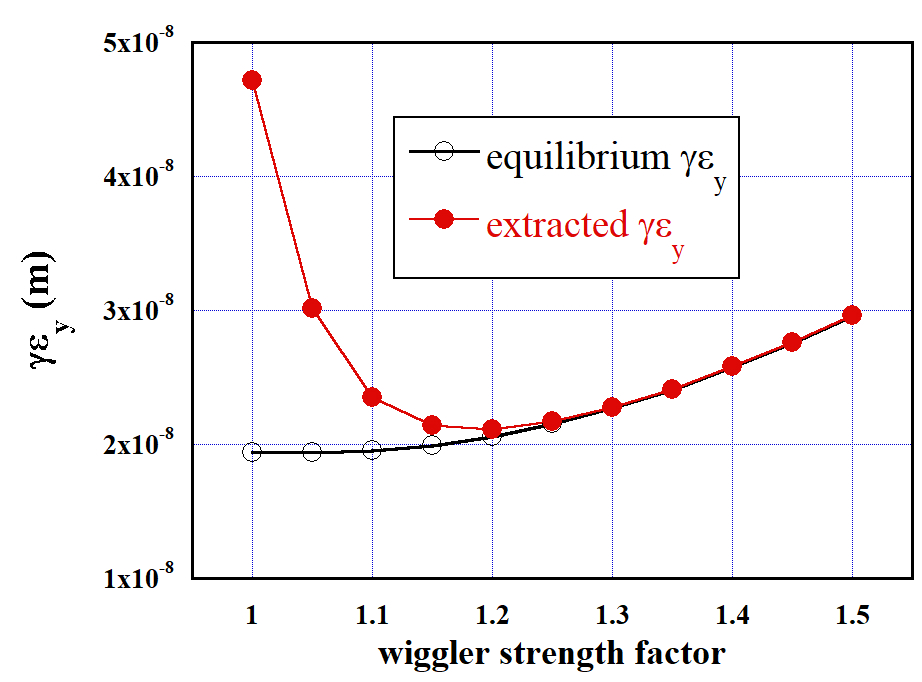}
\end{minipage}
\vspace{-2mm}
\caption{Horizontal (left) and vertical (right) normalized emittance as functions 
of the wiggler field strength. The open circles are the equilibrium emittance 
and the solid (red) circles are the emittance when extracted at 135ms after injection. 
0.005 is assumed as the vertical-horizontal coupling. The effects of the intra-beam 
scattering is included. }
\label{fig:EmittanceFromDR}
\end{figure}

Figure.\ref{fig:EmittanceFromDR} shows the equilibrium and extracted emittances 
from the damping ring as functions of the wiggler strength parameter. The initial 
emittance is $1\times 10^{-3}$ rad$\cdot$m (for positron). As the 
vertical-horizontal coupling, 0.005 is assumed. 

The figure on the right shows that the extracted vertical beam emittance is 
too large if the wiggler strength factor $\lsim$1.1 because the damping is not 
fast enough.  The strength factor $\sim$1.15 ($B=1.48$Tesla) 
gives $\gamma\varepsilon_x \approx 4\mu$m 
$\gamma\varepsilon_y \approx 21$nm. The latter is slightly larger than the 
value in Table \ref{tab:param} (20nm) but a slight improvement of the coupling would be 
sufficient.

Because the wiggler strength is increased, the dynamic aperture of the damping 
rings must be re-examined. The requirements are, as in TDR:
\begin{itemize}
\setlength{\itemsep}{0mm}
\item Normalized betatron amplitude $A_x+A_y \leq 0.07 {\rm m}\cdot{\rm rad}$
\item  Energy deviation $ \leq 0.75$\%
\end{itemize}
The wiggler strength factor is set to 1.15 (curvature 0.0891 m$^{-1}$). 
The acceptance criterion is the survival after 1000 turns. 

\begin{figure}[H]
\begin{minipage}{0.49\textwidth}
\includegraphics[width=\textwidth, clip]{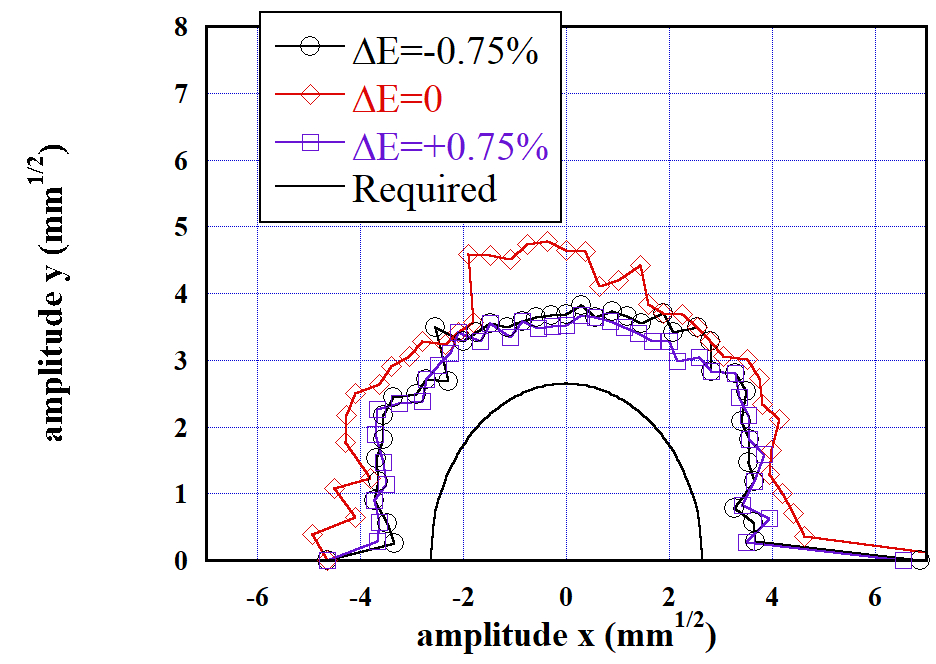}
\end{minipage}
\begin{minipage}{0.49\textwidth}
\includegraphics[width=\textwidth, clip]{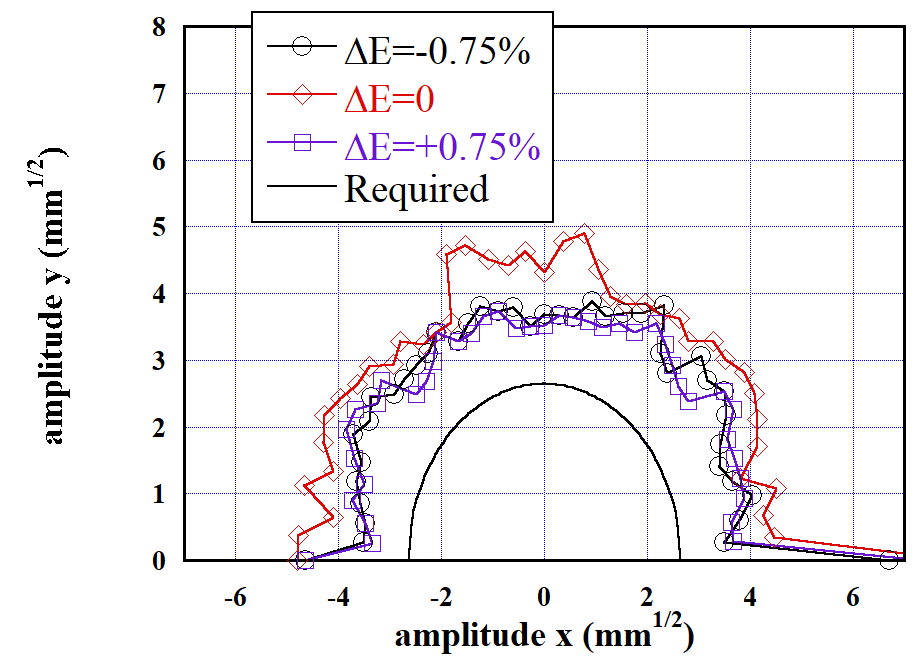}
\end{minipage}
\vspace{-2mm}
\caption{Dynamic aperture of the damping rings with increased wiggler strength. 
The plot on the left (right) is the case without (with) the sextupole components of the 
wiggler field.  
\label{fig:DRdynap}}
\end{figure}

Figure.\ref{fig:DRdynap} show the results. The plot on the left (right) shows the 
aperture without (with) the sextupole components of the wiggler field. 
(Adopted sextupole field is $(\Delta B/B)=0.066$ at $x=10$mm as in TDR\cite{bib:TDR} 
page 115.)  The half circle at the center-bottom shows the required aperture 
for the transverse emittance mentioned above. 
As is seen, the dynamic apeture is sufficient. 
The sextupole field has no significant effects.  
The magnet error has not been included here, but its effect was not significant in the 
previous calculation with the wiggler strength 1\cite{bib:KYmorioka}.

All simulations in this section were done using the computer code SAD\cite{bib:SAD}.

\section{Electron Main Linac}
There are several issues to be studied on the 3.7+3.7Hz operation of the electron 
main linac.
\begin{enumerate}
\item[(a)]  The RF system must deliver different pulses (power and pulse length) 
     alternatingly with an interval of 135ms.
\item[(b)]  The orbits for the 45.6GeV (colliding) and 125GeV (positron production) beams 
      are different even without alignment errors due to the earth's curvature. 
      This difference must be corrected before injection to the undulator section. 
\item[(c)]  The emittance increase by the misalignment must be checked. In partucular 
      the major concern is the colliding beam because the accelerating gradient 
      is very low.
\end{enumerate}
As mentioned in the Sec.\ref{sec:RepRate}, the 45.6GeV (colliding) beam must be 
created by uniform acceleration over the entire linac rather than full gradient 
followed by detuned cavities. There seems to be no other RF-technical problem 
of (a) in the alternating operation. 

\begin{figure}[H]
\begin{center}
\includegraphics[width=0.7\textwidth, clip]{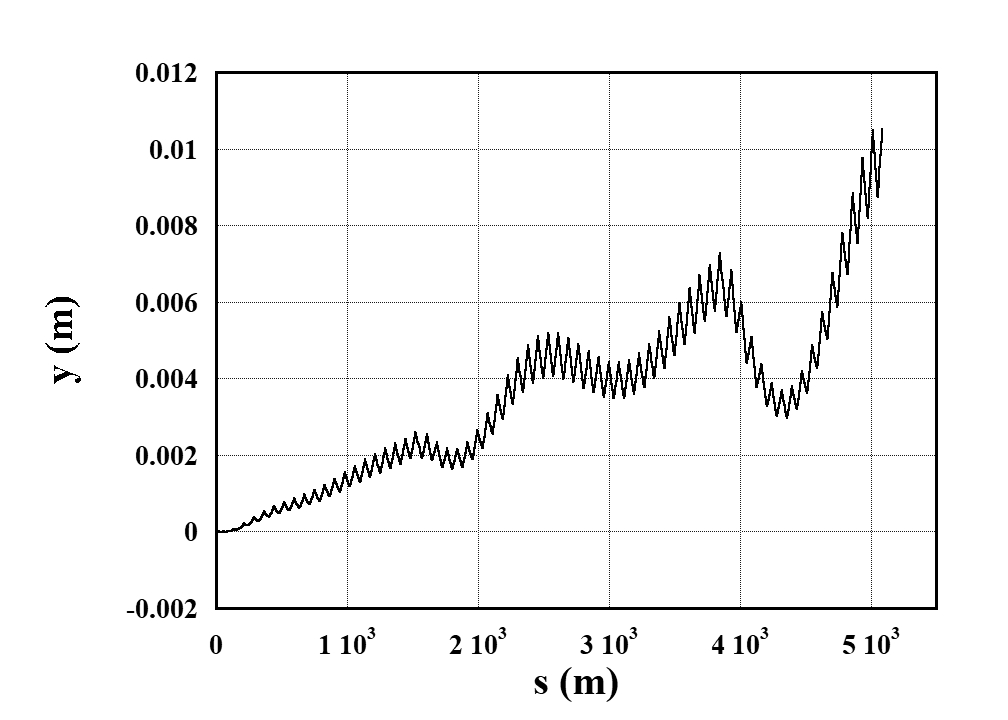}
\end{center}
\vspace{-10mm}
\caption{The vertical orbit difference between 45.6 and 125GeV beams as a function 
of the length $s$ along the linac. No errors included. 
\label{fig:OrbitDiff}}
\end{figure}

A simulation was done for the issue (b). Figure.\ref{fig:OrbitDiff} shows the 
orbit difference. It amounts to $\sim$1cm at the end of the linac. 
This difference (and the orbit angle difference, which is not shown here) must be 
corrected before entering the undulator section by using 3.7Hz pulsed magnets. 
A short section must be allocated (the required length still need to be estimated) 
but there is no problem as far as the field strength is concerned. 
It must be checked whether the field can be accurate enough. In this respect 
it is better to bend the positron production beam, though the energy is higher, 
because the colliding beam must be extremely stable in particular in the vertical plane.

In addition, the wakefield effects (mainly resistive wall) for the colliding beam 
in the undulator section should be studied in more detail because the beam energy 
is low. If the effect turns out to be serious, we have to prepare a beamline 
to bypass the undulator.

For the issue (c) simulations have been done under the following assumptions. 
\begin{itemize}
\setlength{\itemsep}{0mm}
\item The vertically curved orbit as mentioned above.
\item  Orbit correction was done for the colliding beam (45.6GeV). The same 
   steering setting was used for the positron production beam since the emittance 
     increase is not an issue for the latter.
\item The assumed misalignments (every cavity and quadrupole magnet are 
misaligned independently, ignoring the module structure.) 
  \begin{itemize}
  \setlength{\itemsep}{0mm}
  \item Quadrupole magnets offset 0.36mm
  \item Cavity offset 0.67m, tilt 0.3mrad
  \item BPM offset $1\mu$m
  \end{itemize}
  The numbers are r.m.s of Gaussian distribution cut at 3 sigmas.
\item Initial normalized emittance $\gamma\varepsilon_y$=20nm
\item  DFS (Dispersion-free steering) with energy change 20\%.
\end{itemize}
Two cases were simulated for the bunch length and the initial energy spread:
\begin{itemize}
\setlength{\itemsep}{0mm}
\item  Bunch length 0.3mm, energy spread 1.2\%
\item  Bunch length 0.41mm, energy spread 0.9\%
\end{itemize}
The reason of the second case is mentioned in the next section.
In the case of longer bunches the expected emittance increase due to the wake field 
 (combined with misalignment) is larger but that due to the energy spread is smaller.

\begin{figure}[H]
\begin{center}
\includegraphics[width=0.7\textwidth, clip]{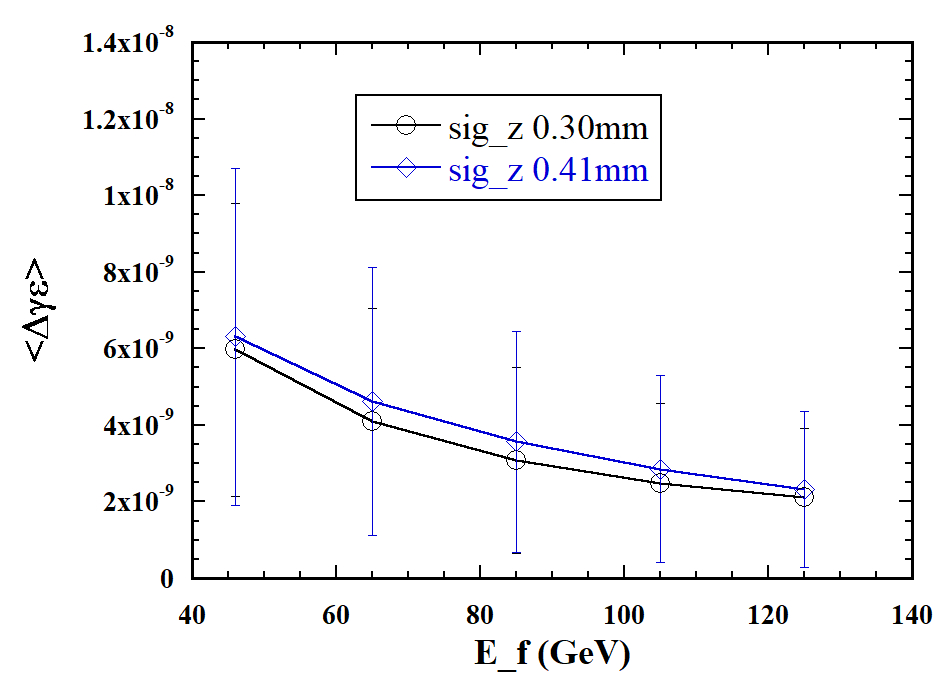}
\end{center}
\vspace{-10mm}
\caption{The vertical emittance increase in the main linacs as a function 
of the final beam energy. Two cases of different $\sigma_z$ are shown.  
The error bars indicate one standard deviation in 100 runs with different 
random number seeds. 
\label{fig:EmitIncrease1}}
\end{figure}

Figure.\ref{fig:EmitIncrease1} shows the vertical emittance increase in the 
misaligned linac as a function of the final beam energy. What is shown is 
the average emittance increase of 100 different random number sets. \footnote{Note: $E_f$=125GeV in this table indicates the case where the orbit tuning is done 
for this energy. This is true for the baseline case $E_{CM}$=250GeV. However, 
the 125GeV beam for the positron production in 3.7+3.7Hz operation 
is different because the tuning is done for 45.6GeV beam. The emittance of 
the positron production beam will be significantly larger. Nontheless, this 
is not an issue because the positron production does not require a very low 
emittance beam.}

The average emittance increase of the 45.6GeV beams with $\sigma_z=$0.41mm 
is 6.3nm, sufficiently smaller than the emittance budget 10nm. 
(Note that the emittance increase would amount to $\sim16$nm for Z-pole 
operation of ILC500\cite{bib:KYmorioka} due to the even lower gradient in longer $\sim$10km linac.) 
The increase in the shorter bunch case is slightly smaller, which indicates 
the effect of the wake is slightly larger than that of the energy spread.

\begin{figure}[H]
\begin{center}
\includegraphics[width=0.6\textwidth, clip]{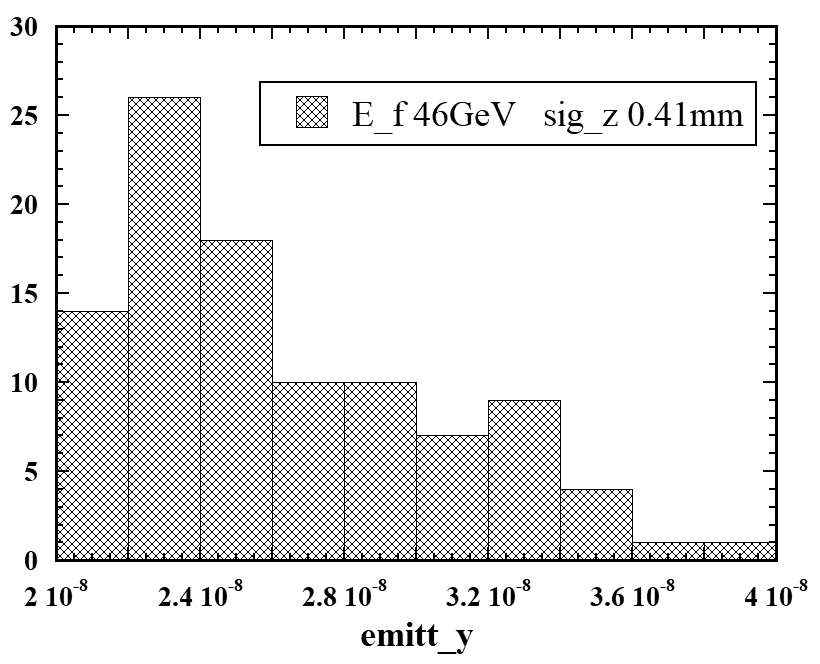}
\end{center}
\vspace{-10mm}
\caption{Distribution of the vertical emittance at the linac end over 100 different 
random number seeds. Final beam energy 45.6GeV. $\sigma_z=$0.41mm.  
\label{fig:EmitIncrease2}}
\end{figure}

Figure.\ref{fig:EmitIncrease2} shows  the distribution of the final emittance  
for the case of the final beam energy 45.6GeV and $\sigma_z=$0.41mm. 
The final emittance at the 90\% confidence level is 
$\gamma\varepsilon_y\sim 33$nm, which slightly exceeds the budget but still 
in an acceptable level.

\section{Beam Delivery System (BDS)}
There are three major issues in BDS related to the Z-pole operation.
\begin{itemize}
\setlength{\itemsep}{0mm}
\item  Collimation depth as explained in the introduction
\item  Momentum band width
\item  Wake field effects
\end{itemize}

\noindent\underline{\bf Collimation Depth}

  The geometric emittance is inversely proportional to the beam energy. 
It is a factor 125/45.6=2.74 times larger than at $E_{CM}$=250GeV. 
The synchrotron radiation from particles in the halo with large angles at the IP 
cause background to the experiment. These particles should be eliminated 
by collimators upstream. However, the required collimation in the horizontal 
plane is too deep ($\sim 6 \sigma_x$) with the TDR parameters even at 
$E_{beam}=125$GeV ($E_{CM}$=250GeV). 
This was the reason that we adopted a smaller horizontal emittance 
$\gamma\varepsilon_x = 10\mu{\rm m} \rightarrow 5\mu{\rm m}$ 
by improving  the damping ring design since 2017\cite{bib:AWLCSLAC}. 
Now, this problem is more serious at the Z-pole. Since the angle at the IP 
is proportional to $1/\beta^*$, to keep the collimation depth 
$\sim 6 \sigma_x$, the horizontal beta function should be 
\begin{equation}
    \beta^*_x \gsim 13{\rm mm} / 2.74 \times (10/5) = \sim 18{\rm mm}
       \label{eq:newbetax}
\end{equation}
where 13mm is the TDR value at $E_{CM}$=250GeV.  

On the other hand there is no problem in the collimation depth in the vertical plane 
because it has a large margin ($\sim 40\sigma_y$) at $E_{CM}$=250GeV. 

\begin{figure}[H]
\begin{minipage}{0.49\textwidth}
\includegraphics[width=\textwidth, clip]{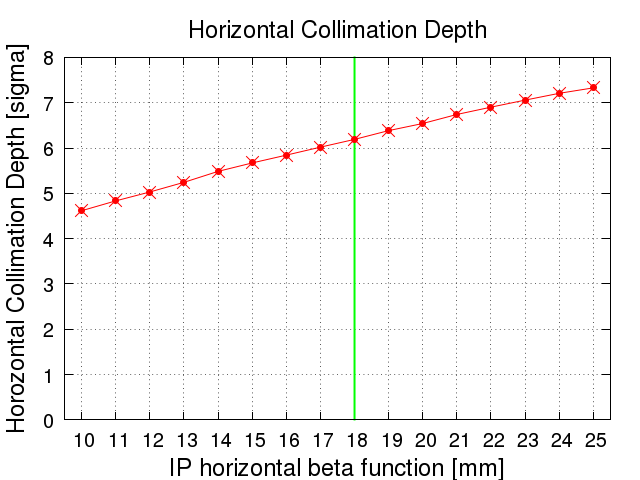}
\end{minipage}
\begin{minipage}{0.49\textwidth}
\includegraphics[width=\textwidth, clip]{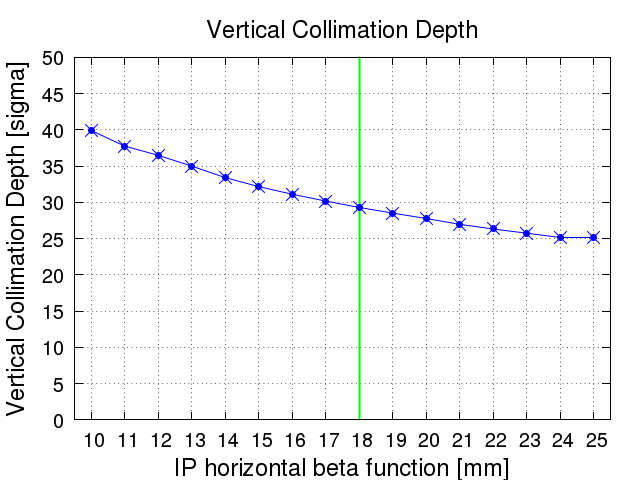}
\end{minipage}
\caption{Collimation depth as a function of the horizontal beta function.
The left(right)-hand-side is the depth in the horizontal (vertical) plane. 
The same beam pipe aperture is used as in ILC 250GeV. 
\protect{$\beta_x^* \times \beta_y^*$} is fixed. 
\label{fig:CollimationDepth}}
\end{figure}

Figure \ref{fig:CollimationDepth} shows the collimation depth obtained 
by simulation. As is seen, the collimation depth in the horizontal plane  
is $\geq 6\sigma_x$ with $\beta_x^*=18$mm, which confirms eq.\ref{eq:newbetax}.

\medskip
\noindent\underline{\bf  Momentum Band Width}

The relative beam energy spread $\sigma_E/E$ at the IP is proportional 
to $1/(E\sigma_z)$. It is $\sim$0.15\% at $E_{CM}$=250GeV\footnote{%
This is the positron energy spread quoted in the TDR. The electron energy spread 
is larger ($\sim$0.19\%) because of the increase in the undulator section. 
Here we use the value for positron for the scaling base because the undulator effect 
is tiny at 45.6GeV.}
If the bunch length (0.3mm) is fixed, the energy spread for Z-pole operation is $\sim$0.41\%. 
This large (relative) energy spread may cause an emittance growth in the BDS 
due to the chromatic effects.

\begin{figure}[H]
\begin{center}
\includegraphics[width=\textwidth, clip]{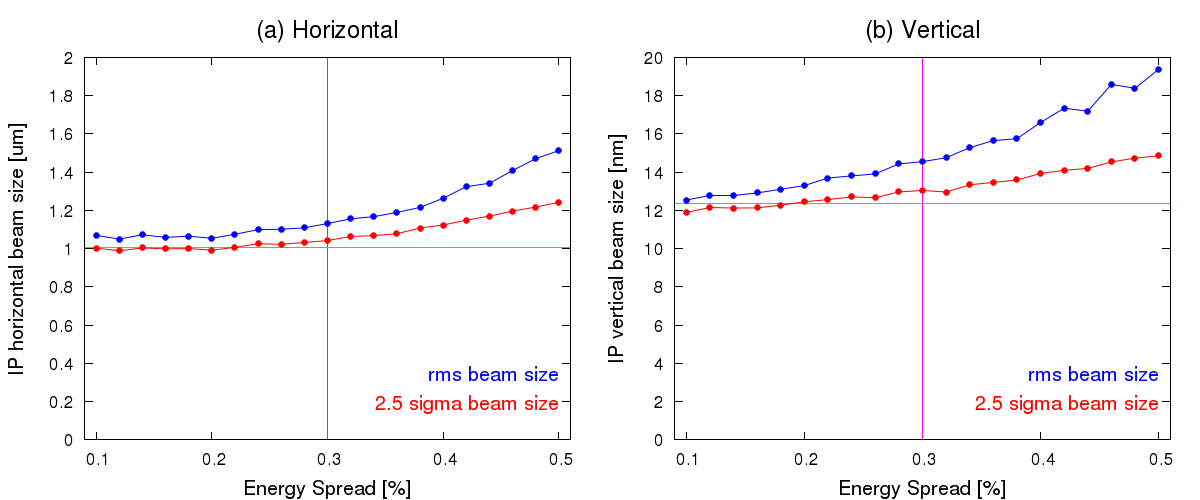}
\end{center}
\vspace{-10mm}
\caption{The beam size at the IP as a function of the energy spread. 
(a) horizontal (b) vertical. The blue spots show the r.m.s.~ of the whole beam size and .
 the red spots show the r.m.s.=size with $>2.5\sigma$ tail cut off. \label{fig:EnergyBW}}
\end{figure}

Figure \ref{fig:EnergyBW} shows the beam sizes as functions of the energy spread. 
Here, the magnet errors are not included. 
This plot shows the beam size growth is too large if the energy spread is 0.41\%. 

To reduce the energy spread, we adopt a longer bunch length in the bunch 
compressor 0.3 $\rightarrow$ 0.41mm, which will bring about a smaller 
energy spread $\sigma_E/E = 0.41{\rm \%} \rightarrow 0.3{\rm \%}$ 
at the IP. 
The beam sizes at IP (2.5$\sigma$) in such a case are 
$\sim 1.04\mu$m (horizintal) and $\sim 12.7$nm (vertical) in the absense 
of magnet errors and wakefields from Figure \ref{fig:EnergyBW}. 

This long bunch has two side effects.

First, the transverse wake field in the main linac and the BDS increases by the factor 
0.41/0.3. The previous section showed that the emittance increase in the main 
linac due to this reason is acceptable.

Another side effect is the increased disruption parameter at the IP. 
This will be mentioned in the next section.

\medskip
\noindent\underline{\bf Tuning of the BDS with magnet errors and wakefield}

Simulations including the magnet errors and the process of beamline tuning 
were done, assuming the errors listed in Table \ref{tab:BDSmagerrors}. 

\begin{table}[H]
\caption{Magnet errors used in the simulation of BDS
\label{tab:BDSmagerrors} }
\begin{center}
\begin{tabular}{|l|l|rl|}
\hline
\multirow{4}{16mm}{Bend}  & rotation   &  100  &  $\mu$rad    \\
               &    field strength    ($\Delta\rho/\rho$)    &    $1\times 10^{-4}$     &     \\
               &    bend-to-BPM alignment    &   100   &  $\mu$m  \\
\hline
\multirow{4}{16mm}{Quad}  &  alignment ($x$,$y$)  &    100   &  $\mu$m  \\
              &   rotation       &  100  &  $\mu$rad    \\
               &    field strength   ($\Delta K_1/K_1$)    &    $1\times 10^{-4}$     &     \\
          & sextupole component ($B_2/B_1$ at $r$=1cm)  &     $1\times 10^{-4}$  &   \\
               &    quad-to-BPM alignment    &  5   &  $\mu$m  \\
\hline
\multirow{4}{16mm}{Sextupole}  &  alignment ($x$,$y$)  &    100   &  $\mu$m  \\   
      & rotation   &  100  &  $\mu$rad    \\
      &    field strength   ($\Delta K_2/K_2$)    &    $1\times 10^{-4}$     &     \\              
      &    sext-to-BPM alignment    &  5   &  $\mu$m  \\
\hline
\end{tabular}
\end{center}
\end{table}

\begin{figure}[H]
\begin{center}
\includegraphics[width=\textwidth, clip]{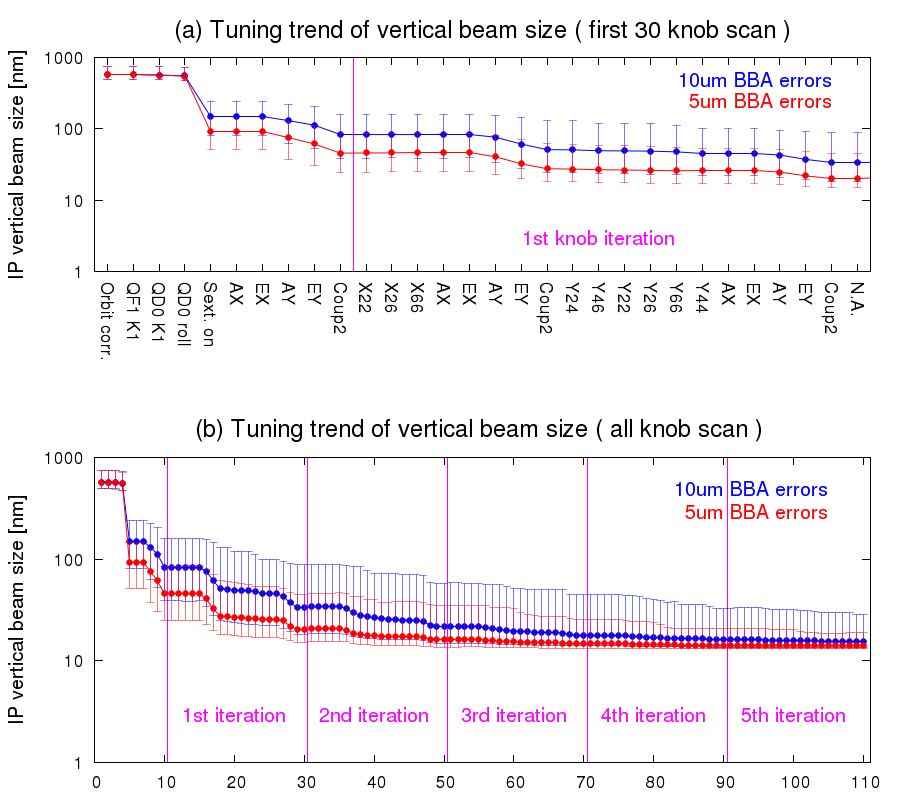}
\end{center}
\vspace{-5mm}
\caption{Vertical beam size vs.~number of knob scans in the process of BDS tuning. 
(a) shows the first 30 knobs. (b) shows all knobs.  \label{fig:BDStuningprocess}}
\end{figure}

Figure \ref{fig:BDStuningprocess} shows a example of the tuning process 
in the presense of magnet errors (but no wake fields). 
It shows the vertical beam size vs.~number of tuning knobs. 
Two cases of different r.m.s.~errors of the distance from the BPM center 
to the quadrupole/sextupole field center (5$\mu$m and 10$\mu$m) are shown.   
10$\mu$m is the value used in the BDS simulations of ILC so far. 
Improvement to $\sim 5\mu$m during a few years of operation experience will be possible. 

Table \ref{tab:BDSbeamsize} shows the beam sizes obtained in this way. 
These values are averages of 100 random number seeds. 
The case with magnet errors and tuning is shown in the row (2).

\begin{table}[H]
\begin{center}
\caption\protect{Simulation results of the vertical beam sizes. \\ 
{\small The offset 300$\mu$m of the wake sources is used for the row (3) and (4). }  
\label{tab:BDSbeamsize}}
\begin{tabular}{|l|rr|}
\hline
                    &   $\sigma_x^*$  ($\mu$m)   &   $\sigma_y^*$ (nm)   \\
\hline
  (1) No errors     &   1.04    &    12.7   \\
  (2) Magnet errors and correction &     1.12    &   14.0       \\
  (3) Magnet errors + static wake + correction  &            &   14.3   \\
  (4) Magnet errors + static\&dynamic wake + correction   &    &   14.6 \\
\hline
\end{tabular}
\end{center}
\end{table}

Since the beam energy is very low at the Z-pole compared with the baseline energy 
$E_{CM}=250$GeV and, moreover, we adopted the longer bunch, the effects of the 
transverse wake are expected to be significant. 

The most important wake sources are the 107 BPMs (Beam Position Monitor) 
distributed over the BDS beamline. They are associated with bellows, and flange gaps. 
We assumed the wake sources are dislocated by 300$\mu$m in the simulations. 

The correction of the wake effects is done by adding artificial wake sources mounted 
on movers  and by changing their position so as to cancell the wake effects. 
This is done only in the vertical plane since the effects on the horizontal plane is 
minor. The row (3) in Table \ref{tab:BDSbeamsize} the result with static wake effects 
and tuning. In this simulation we assumed that the wakefields 
by bellows and flange gaps are minimized by using RF contacts. If not, the 
resulting beam size would be $\sim$15.8nm instead of 14.3nm. 

In addition to the static wakefield effects described above, 
there is a `dynamic effects' coming from the beam jitter from upstream. 
Since many wake sources are placed where $\beta$ is large and the phase advance 
 from the IP is $\pi/2$, the beam jitter can be enhanced by the wakefields.  
It is compensated for by the fast feedback system developed for the ILC. 
From our experience we assumed  the feedback system can suppress the 
beam angle jitter to 10\%. 

The row (4)  in Table \ref{tab:BDSbeamsize} shows the vertical beam size in 
the presense of the static/dynamic wake effects and the correction. 
The resulting vertical beam size was 14.6nm. This is adopted in Table \ref{tab:param}. 

These simulations show that the following issues are important for the BDS 
tuning especially at the Z-pole.
\begin{itemize}
\item  Accurate alignment between the magnet field centers and the electric 
center of BPMs.
\item  Sufficient RF contact for the bellows and flange gaps to minimize the wakes.
\item  Relaxing the intensity dependence due to the static wake effects by 
         wakefield knobs.
\end{itemize}

\section{Beam-Beam Interaction}

\begin{figure}[t]
\begin{center}
\includegraphics[width=0.9\textwidth, clip]{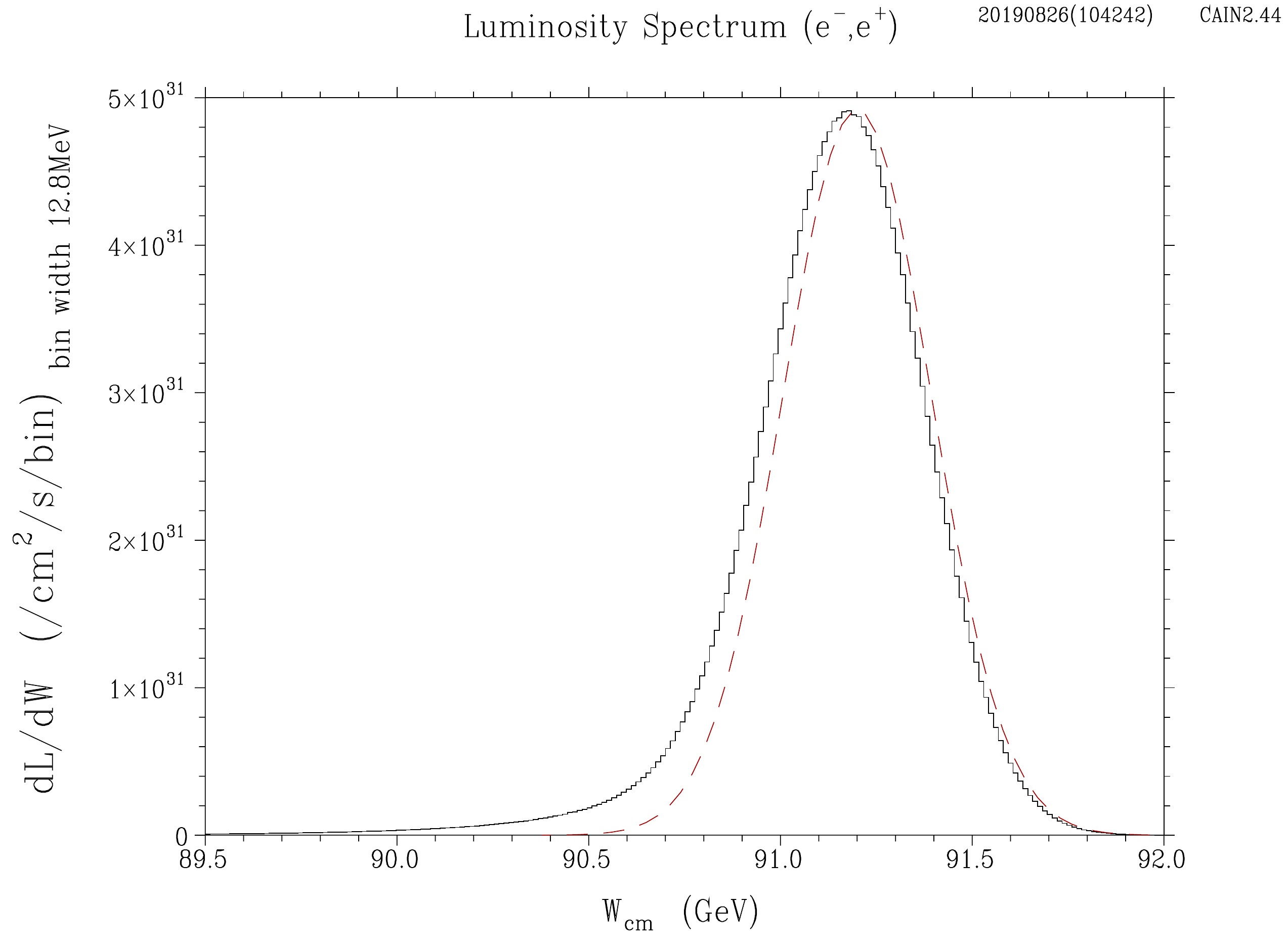}
\end{center}
\vspace{-10mm}
\caption{Luminosity spectrum. The red dashed curve shows the initial distribution normalized 
at the peak. 
\label{fig:LuminositySpectrum}}
\end{figure}

Figure.\ref{fig:LuminositySpectrum} shows the luminosity spectrum with the 
parameters listed in Table \ref{tab:param}. The dashed curve shows the initial 
distribution normalized at the peak. As is seen, the contibution of the 
beamtrahlung is very small. 

Because of the longer bunch to reduce the energy spread, the vertical disruption 
parameter is slightly enlarged ($D_y=31.8$). 
Nonetheless, it is still lower than the value 34.5 for $E_{CM}=125$GeV case. 
Therefore, the tolerance of the beam offset will not be severer ($~0.1\sigma^*_y$).

\section{Luminosity Upgrade}
In the present design the major limitation of the luminosity comes from the 
energy band-width of the final focus system and the large vertical disruption parameter. 

The former may be a little improved by adopting a shorter final quadrupole magnet. 
In fact the TDR adopted 2m long magnet which was split into two 1m magnets.  
Both are excited for $E_{CM}=500$GeV but the upstream one is turned off for  
low energy operations. This helps in increasing the horizontal collimation depth from 
$\sim 5\sigma_x$ to $\sim 6\sigma_x$ at $E_{CM}=250$GeV. 
However, we cannot expect a large improvement of the collimation depth 
at the Z-pole operation by a further reduction of the magnet length, say, to 50cm. 

The latter is an issue of the beam position control at the IP. 
Since the beamstrahlung is negligible at Z-pole, one can futher reduce the horizontal beam 
size in this respect,  if an even larger disruption parameter is acceptable. 
The adoption of the so-called traveling focusing might allow a larger disruption 
but this is beyond the scope of this note. 

On the other hand we do not forsee a major obstacle in doubling the luminosity 
by doubling the number of bunches from 1312 to 2625. Once the RF system 
is reinforced for doubled bunches, we expect to reach 
${\cal L} \sim 4.2\times 10^{33}/{\rm cm}^2/{\rm s}$ at Z-pole.

 When ILC is extended to 500GeV, there will be a chance to adopt the scheme 
 proposed in \cite{bib:NWalker}, where the 250GeV electron main linac is split into 
 two parts, upstream part ($>80$GeV for W-pair) for the colliding beam and 
 the downstream part ($>125$GeV with an additional electron gun) for positron production. 
 If this possibility is to be taken into account later, some preparation is necessary 
 for the first construction (e.g., a space for the transport line of the colliding beam).  
 
\section{Summary}
We estimated the luminosity of ILC for 250GeV at the center-of-mass energy 
for Z-pole. In addition to the change of the horizontal emittance, which has already 
been announced for maximizing the luminosity at 250GeV, 
we made the following changes of the parameters:
\begin{itemize}
\setlength{\itemsep}{0mm}
\item  Reduced the repetition rate from 5+5Hz to 3.7+3.7Hz because of the 
         limited power consumption of the electron linac.
\item  Increased the bunch length from 0.3mm in TDR to 0.41mm in order to 
     reduce the beam energy spread for relaxing the issue of the momentum bandwidth 
     of the final focus system.
\item  Increased the $\beta^*_x$ from 13mm for $E_{CM}=250$GeV to 18mm 
     for relaxing the requirement of the horizontal collimation.
\end{itemize}
The resulting luminosity is  $2.1 \times 10^{33}/{\rm cm}^2/{\rm s}$, 
which is significantly higher than the value ($\sim 1.5 \times 10^{33}$)  from the 
previous simple scaling in spite of the reduced repitition rate. 
$\propto E^{1.5}$ from 250GeV in TDR. 

It is presumably possible to double the luminosity when the doubled bunch option 
(2625 bunches) of ILC is realized.

\end{document}